# Exciton emissions in bilayer WSe$_2$ tuned by the ferroelectric polymer

Sixin Zhu[†#], Dan Li[ξ], Jianlu Wang[ξ], Qiang Wang[†#], Yongpeng Wu[†], Liang Xiong[†], Zhanfeng Jiang[†], Huihong Lin[#], Zhirui Gong[†*], Qi Qin[†*], Xingjun Wang[ξ*]

[†]College of Physics and Optoelectronic Engineering，Shenzhen University, Shenzhen, 518060, People's Republic of China

[#]School of Chemical and Environmental Engineering, Hanshan Normal University, Chaozhou, 521041, Guangdong, People's Republic of China

[ξ]State Key Laboratory for Infrared Physics, Shanghai Institute of Technical Physics, Chinese Academy of Sciences, Shanghai, 200083, People's Republic of China

**ABSTRACT:**

In this work, a hybrid integration of few-layer transition metal dichalcogenides (TMDCs) and ferroelectric polymer is designed to achieve passive control of optical properties in-situ. The electrical polarization in ferroelectric P(VDF-TrFE) polymer can regulate the photoluminescence (PL) in few-layer TMDCs. The total PL intensity is substantially suppressed or enhanced under opposite polarization in bilayer $WSe_2$. This is because electrons transfer between valley K and Λ in the conduction band induced by the built-in electric field in P(VDF-TrFE) polymer. This charge transfer further changes the competing dynamics between direct and indirect exciton recombination path and overall optical radiation efficiency. We also illustrate that the engineered PL originates from external electric field dependent transferred electron effect. The theoretical result matches the experimental data well. This work demonstrates a device platform in which passive regulation is achieved using 2D TMDCs modulated by polarized ferroelectric materials.

In the past decade, 2D materials have been widely studied as active materials in optoelectronic devices such as photodetectors,[1,2] photovoltaic cells,[3-5] or light emitters [6-8] etc. However, these devices are usually required to be continuously electrically biased in order to function properly, which often leads to relatively significant power consumption. For achieving low power operation, an alternative method is to utilize non-volatilitous electrostatic fields of ferroelectric crystal at the interface. In recent years, 2D multilayer materials/P(VDF-TrFE) polymer hybrid structure is considered as a promising optoelectronic device platform.[9-11] Poly (vinylidene fluoride-trifluoroethylene) (P(VDF-TrFE)) polymer is a flexible and transparent ferroelectric material, which can be conveniently prepared by spin coating at low temperature.[12,13] In the meantime, multilayer TMDCs have high electron mobility and large optical density of state. In the hybrid device, P(VDF-TrFE) polymer can provide an ultra-high local electrostatic field (~ 1 V/nm) in adjacent thin-film semiconductor materials. This electrostatic field is larger than counterpart induced by gate bias in traditional field effect transistors (FETs). Furthermore, they can work as a self-maintaining devices without externally bias voltage. 2D multilayer materials/P(VDF-TrFE) polymer hybrid structure has many unique properties that do not exist in traditional FETs or p-n junctions. For example, a photodetector based on $MoS_2$/P(VDF-TrFE) hybrid structure has response up to the near-infrared,[14] while devices with $MoS_2$/non-ferroelectric materials, such as $SiO_2$, can only respond in visible range. In addition, $MoTe_2$ p-n junction is realized using the altered nonvolatile electric field induced by the P(VDF-TrFE) poled in the opposite direction, which can respond up to 1400 nm at room temperature.[15]

Extensive efforts have been devoted to engineering the optical properties of ultrathin TMDCs using vertical electric field. In this configuration, it is expected that TMDCs' properties, such as bandgap and photoluminencence (PL), are modulated in a non-volatile and reversible fashion, while carrier mobility in TMDCs is kept. First, the bandgap engineering of ultrathin TMDCs have been studied. The bandgap of ultrathin TMDCs decreases monotonously when the vertical electric field is increased owing to the giant Stark effect, as predicted by density functional theory (DFT) calculations.[16,17] However, in either top-gate or bottom-gate configurations based on traditional non-ferroelectric gate FETs, no significant exciton energy change in TMDCs is experimentally observed. This is because that the electric field in this case only adjusts TMDCs' Fermi energy.[18] In recent year, using dual-gated field-effect transistors, significant bandgap modulation in multilayer TMDCs has been experimentally realized.[19-23] In this configuration, Fermi energy and electrical displacement field can be controlled by the dual-gates independently. Secondly, PL modulation of ultrathin TMDCs has been explored. In monolayer TMDCs, the PL intensity is modulated by vertical electric fields because the free electron or hole concentrations changes when Fermi energy is adjusted by the electric field.[18,24-26] In bilayer TMDCs, PL from inter-layer exciton is quenched under the vertical electric field in the dual-gate FET structure, but the intralayer excitons has no corresponding electric field response.[20,23] However, engineering the intralayer excitons emission of few-layer TMDCs in-situ under the vertical electric field is highly desired for optoeletronic devices. In view of the particularity of the hybrid structure of ferroelectric materials and 2D TMDCs, we are dedicated to studying the regulation of the optical properties of few-layer TMDCs by ferroelectric materials, while the PL modulation of multilayer TMDCs in-situ using ferroelectric materials has not yet been relevant reported.

In this work, we demonstrate that the PL intensity of intralayer excitons in bilayer WSe$_2$ can be modulated (suppressed or enhanced) reversibly utilizing bilayer WSe$_2$/ P(VDF-TrFE) polymer structure. The electrical polarization of P(VDF-TrFE) polymer is regulated by an external gate voltage. The experimental PL intensity is substantially suppressed or enhanced under opposite ferroelectric polarization due to electrons hop between the conduction band energy valley K and Λ under the action of the P(VDF-TrFE) electrostatic field. This further change carrier recombination path and overall optical radiation efficiency. Experimental results are matching with the theoretical results based on transferred electron effect. This work may pave the way for electrically tunable photoluminescence in multilayer TMDCs.

**RESULTS AND DISCUSSION**

The bilayer WSe$_2$ samples were mechanically exfoliated from bulk WSe$_2$ onto SiO$_2$ (300 nm)/Si substrates. The device has a two-terminal configuration which is illustrated in Figure 1a. The P(VDF-TrFE) polymer is the top-gate dielectric material and has three polarization states: no polarization, polarization up state (P+) and polarization down state (P-). The P+ and P- states are achieved by poling P(VDF-TrFE) polymer with negative top-gate voltage (-V$_{tg}$) and positive top-gate voltage (+V$_{tg}$) (see Figure 1b). The ferroelectric hysteresis loop of 300 nm P(VDF-TrFE) at room temperature is shown in Figure 1c. The polarization saturates when the voltage approaches ± 60 V, beyond which the polarization strength increases slowly as the voltage further increases. The coercive voltage is about 22.5 V and the remnant polarization value is approximately 7 μC cm$^{-2}$. Figure 1d shows the typical PL spectrum after coating P(VDF-TrFE) polymer on the bilayer WSe$_2$. The direct and indirect exciton emission in bilayer WSe$_2$ are extracted by Gaussian fitting method, where the PL peak (A) at 1.58 eV originates from the direct interband transition and the peak (I) comes from the indirect bandgap emission.

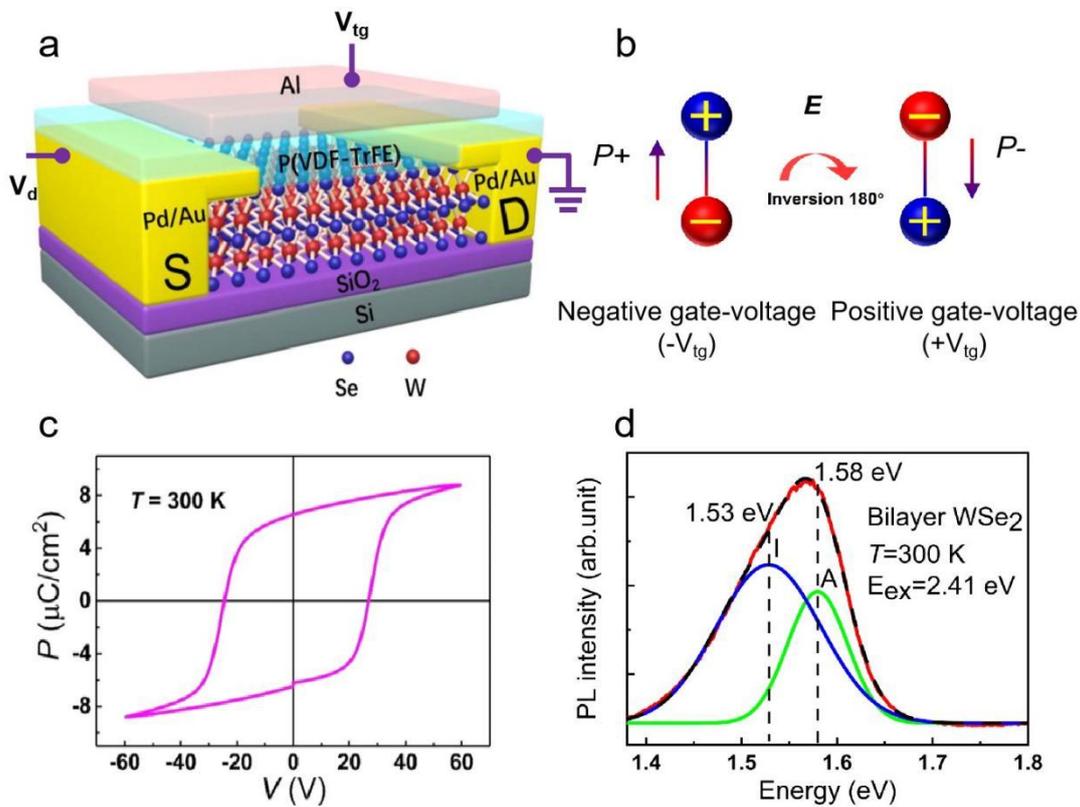

**Figure 1**. (a) A 3D schematic diagram of the two-terminal device configuration after coating with P(VDF-TrFE). The Pd/Au (15 nm/45 nm) serves as the drain and source electrodes and Al (10 nm) serves as the top-gate electrode. (b) The polarization up state and polarization down state are achieved by poling P(VDF-TrFE) polymer with negative top-gate voltage (-V$_{tg}$) and positive top-gate voltage. (c) The ferroelectric

hysteresis loop of 300 nm P(VDF-TrFE) at room temperature. (d) Bilayer WSe$_2$ PL spectra at room temperature. The blue and green lines indicate the direct (A) and indirect (I) interband transitions.

The exciton emission of bilayer WSe$_2$ is significantly regulated by the P(VDF-TrFE) polarzation state. The experimental PL intensity is substantially enhanced/suppressed with opposite continuous polarization (the external gate-voltage always exists) as illustrated in Figure 2a and 2b, and the direction of the arrows in the figure indicate the direction in which the polarization intensity increases. By properly controlling the experimental conditions, the PL intensity can be adjusted (enhanced or suppressed) approximately 3 times. In addition to the intensity modification, the vertical electric field also causes a slight blue-shift (corresponding to P+) or red-shift (corresponding to P-) of the highest peak of the characteristic PL. The Insets in Figure 2a and 2b show the relationship between the total integrated PL intensity and the top-gate voltage. As the top-gate voltage reach a certain threshold value (~ ±60 V), the total PL intensity saturates and increases slowly. These spectral characteristics match the trend of the ferroelectric hysteresis loop well. Therefore, the changed PL instensity is temporarily attributed to the regulation of P(VDF-TrFE) polarization, which is further confirmed by our following experimental and theoretical analysis.

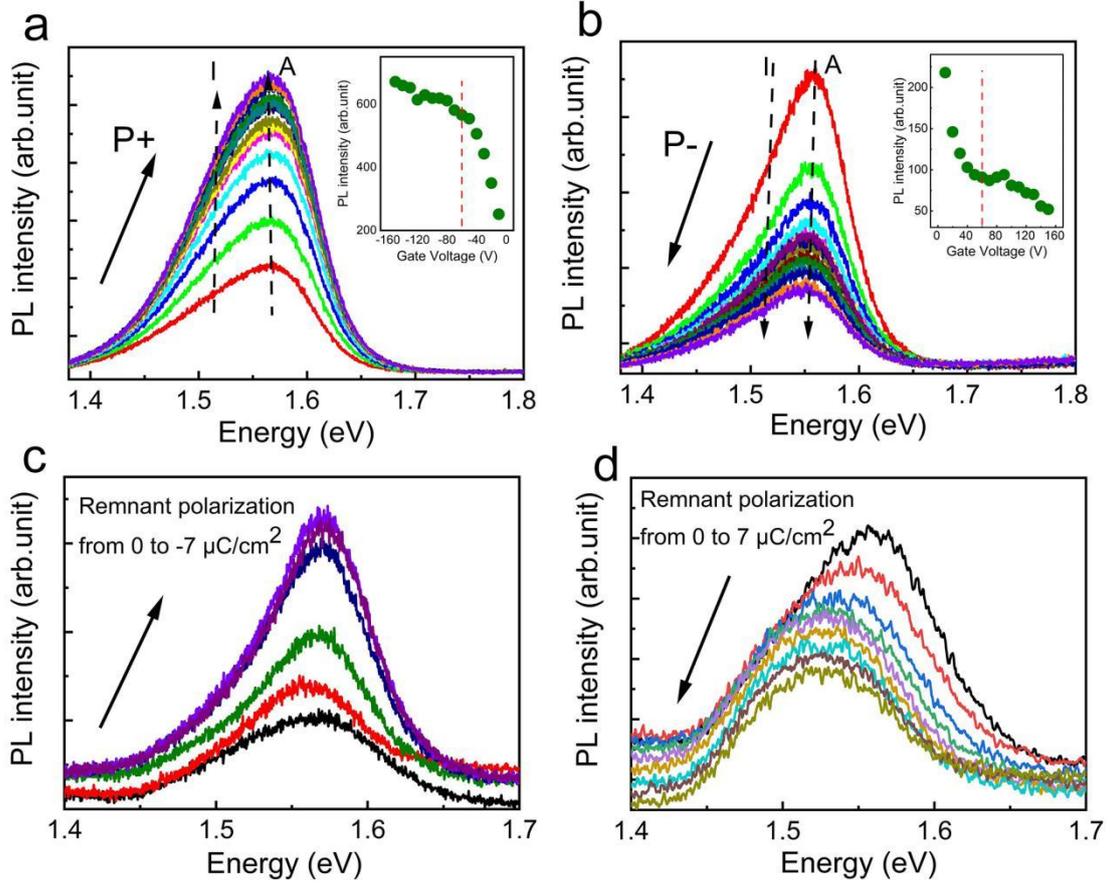

**Figure 2**. (a, b) PL spectra at continuous polarization up state (P+) and polarization down state (P-) with bilayer WSe$_2$. The change of PL intensities with the increase of the applied top-gate bias is show in the inset. (c, d) PL spectra of bilayer WSe$_2$ with P(VDF-TrFE) remnant polarization. The applied gate voltage is removed after poling P(VDF-TrFE) with different top-gate voltages. The direction of the arrows in the figure indicate the direction in which the polarization intensity increases.

To further confirm the mechanism of this PL intensity modulation, we also conduct PL measurements when bilayer WSe$_2$ is regulated by the remnant polarization of P(VDF-TrFE) polymer. It should be noticed that in Figure 2a and 2b, V$_{tg}$ is always kept on, while in Figure 2c and 2d the applied gate voltage is removed after P(VDF-TrFE) is polarized with different gate voltages. The total PL intensities increase or decrease with remnant polarization (Figure 2c and 2d), and then remains almost constant after reaching a critical value (polarization~ -7 μC cm$^{-2}$ or above $V_{tg}$ ~ -60 V). In addition, the highest peak of the characteristic PL exhibits a signification shift under the action of remnant polarization. This observed experimental spectrum characteristics is substantially consistent with the Figure 2a and 2b when $V_{tg}$ is alway kept on. After $V_{tg}$ is removed, only the remannent polarization of P(VDF-TrFE) is changed compared to the original state before $V_{tg}$ is applied. This is the dominant factor to modulate the PL intensity. The deformation of P(VDF-TrFE) thin film caused by the inverse piezoelectric effect can also be excluded. In addition, the bias voltage directly applied to bilayer WSe$_2$ basically has no modulating effect on the PL intensity ( in Figure S2). Therefore, it can be concluded that the changed PL intensity is due to the electric field in the vicinity of polarized P(VDF-TrFE) polymer.

Previously reported works propose some mechanisms to accounted for the PL intensity variation under the action of an electric field, such as thermal effects,[27] exciton ionization,[28] spatial separation of electrons and holes [20] and the intervally charge transfer [29], they can not fully explain this experimental results. The increase in gate voltage may also cause sample to heat up. Temperature may decouple adjacent layers through interlayer thermal expansion or enhance the electron-phonon interaction, leading to an increasing or decreasing in PL intensity. Combined with the temperature-dependent PL measurements (see Figure S3 in the Supporting Information), the thermal effect can be eliminated because the PL quenching processes (such as Auger decay) is dominant. In addition, when the kinetic energy of charged particles is large enough, exciton ionization may occur through the collision process, thereby changing the PL recombination pathway. However, this ionization process only leads to the quenching of exciton PL intensity. For spatial separation of electrons and holes, the vertical electric field in dual-gate bilayer TMDCs FETs will only cause the PL intensity of interlayer excitons to be quenched, due to localize electrons and holes in different layers and reduce the overlap between them, where the "intralayer exctions" do not respond to the vertical electric field.

While the above analysis confirms that the PL intensity is regulated by PVDF polarization, the following study attributes microscopically the changed PL intensity is because of the transfer of photoexcited electrons between the conduction band extremum. In semiconductor materials, electrons can transfer from one conduction band extremum to another facilitated by an electric field.[30] According to the opposite changing trends of the highest peak of PL under different polarization direction, the change in the intralayer exciton emission intensity is first tentatively attributed to the change in the number of photogenerated carriers at the extremes of conduction band. This further changes the dominant radiation recombination path. In order to confirm this mechanism, we conduct electronic structure calculation using density function theory (DFT). Then using the transferred electron theoretical model, we numerically extract some key parameters and compare the extracted data with experimental result to verify the proposed model.

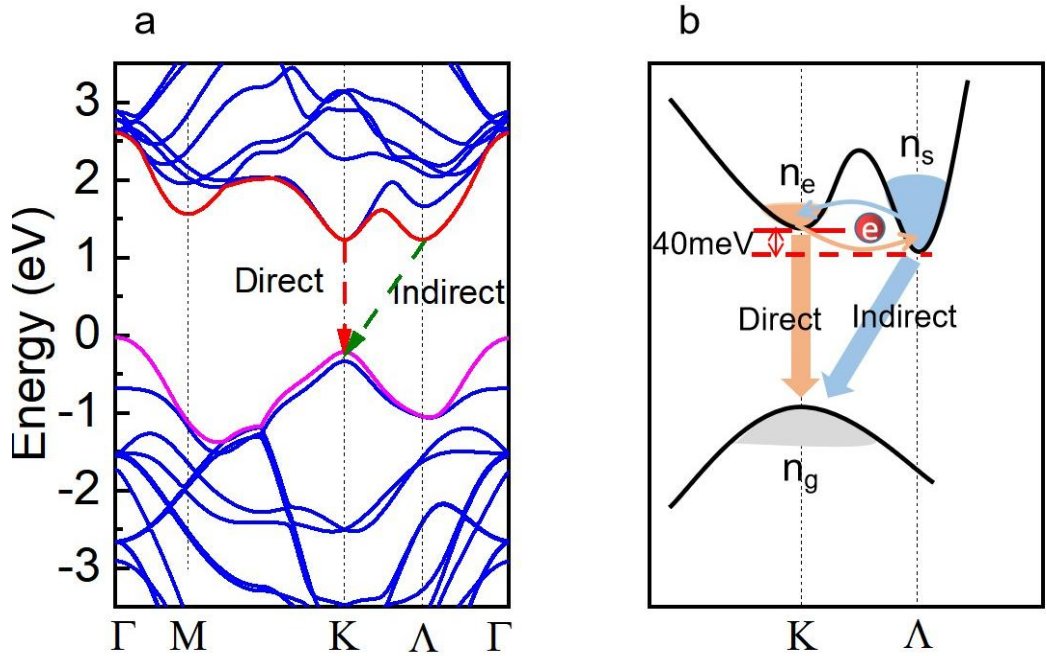

**Figure 3**. (a) Calculated energy band structure of the bilayer $WSe_2$. The dashed arrows imply the possible radiation recombination pathways. (b) Electron transfer effect under strong electric field.

According to the DFT calculation (Figure 3a), the indirect band emission involves holes at the K point of valence band holes and electrons at Λ point of conduction band, while the indirect transition happens between K point of conduction and valance band.[31,32] In the conduction band, the energy difference between the K and Λ point is about 50 meV, which matches the PL peak distance to the experimental value (~ 50 meV). Because of the small energy difference between the K and Λ point of conduction band, electron can transfer between different energy valleys relatively easily facilitated by phonon (momentum conversation) and applied electric field (Figure 3b). According to the quasi-Boltzmann distribution law, the number of photogenerated electrons at Λ point dominates when there is no external perturbation.

When the P(VDF-TrFE) polymer is in P+ polarization, electrons transit from Λ point to K point in bilayer WSe$_2$. The electron population ratio between K piont and Λ point is increased. On the other hand, when the P(VDF-TrFE) polymer is in P- polarization, electrons transfer from K point to Λ point. Since the quantum efficiency of the direct transition is much higher than the indirect transition, the increase in the number of electrons in the K energy valley will inevitably lead to an increase in the entire fluorescence intensity, and vice versa. In addition, the direct exciton and indirect exciton emission intensity are extracted by Gaussian fitting method (see Figure 1d). As shown in Figure 4b, the PL intensity ratio of direct and indirect excitons decrease gradually with the applied bias change from -150 V to +150 V. With increasing the applied negative bias voltage, the Λ valley electrons will transfer to K valley via electric-assisted intervally scattering effect. As a result, the carrier population at K point is increased and the direct exciton radiation transition is enhanced. This result is also consistent with the aforementioned intervallery electron transfer model. Qualitatively, this matches the experimental hysteria.

To further verify the theoretical model, the qualitative discussion is as follows. The PL intensity of both the direct and indirect excitons are proportional to the product of three factors - corresponding populations $n$, transition probability $A$ and energy difference $\Delta$ ( $I_{PL} \propto nA\Delta$ ). The transition probability $A$ is kept same because it is determined by a constant oscillator strength in this study. Therefore, the PL intensity mainly depends on the exciton population. Then we need to establish the relationship between ferroelectric polarization/gate voltage and the exciton population.

When an external electric field is applied on a completely unpolarized ferroelectric material, the electronic polarization of the ferroelectric material between the electric dipole moment $P$ and external electric field $\xi$ is approximately expressed by

$$P = -P_0 \tanh\left(\frac{\xi}{\xi_0}\right) \quad (1)$$

where $P_0$ is the maximum polarization and $\xi_0$ denotes the hardness of the ferroelectric material. By solving the rate equation of the exciton populations of the direct and indirect exciton states (see supplementary material for further details), the population of the effective direct exciton state proportional to the PL peak intensity is given as

$$n_e(\xi) \approx \beta \exp\left(-\frac{\delta+\Delta}{k_B T}\right)\frac{R(P)(1+r)}{R(P)r+1} \quad (2)$$

where $\delta$ is the energy difference of the conduction band extreme between K point and $\Lambda$ point, $\Delta$ is the energy gap of the bilayer WSe$_2$, $k_B$ is Boltzmann constant and $T$ is the temperature. Here, $r = \beta g_e/g_s \cdot \exp(\delta/k_B T)$ where $g_s$ and $g_e$ are the excitation rates of the indirect exciton and direct exciton, $\beta$ is the oscillating strength ratio between the direct exciton and the indirect exciton. According to the transferred charge effect based on the effective electron temperature,[30,33] the external field dependent ratio $R(P)$ can be expressed as

$$R(P) = \exp(-k \tanh\left[\frac{\xi}{\xi_0}\right]) \quad (3)$$

with the dimensionless parameter

$$k = \frac{2}{3}\frac{\delta}{\left(k_{B}T\right)^{2}}\frac{pP_{0}}{\varepsilon} \qquad (4)$$

$p$ is the electric dipole and $\varepsilon$ is the dielectric constant of the P(VDF-TrFE) polymer.

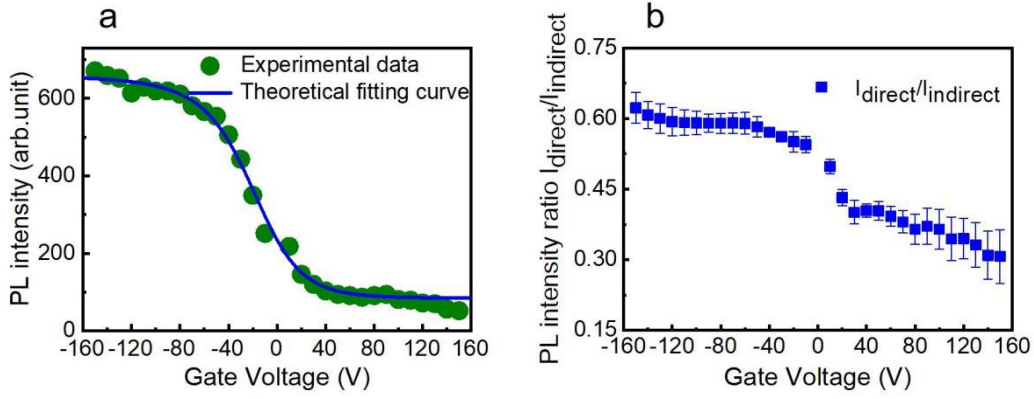

**Figure 4**. (a) The PL intensity of bilayer $WSe_2$ at different gate voltage. Circle represents experimental data. Solid lines show the fitting PL intensity curves using the equation 2. (b) The PL intensity ratio of direct and indirect exciton at bilayer $WSe_2$.

Figure 4a shows the total integrated PL intensity of direct and indirect excitons as a function of the gate voltage. The calculated PL intensities indicated with the blue solid lines (by fitting the Eq.2 with the experimental data), which are well matched with the experimental results. The fitting parameters are $k = 1.24$, $r = 0.11$ and $\xi_0 = 46.76$ V for $\delta = 50$ meV, $\Delta = 1.53$ eV and $k_B T \approx 25.9$ meV for the room temperature. From the above analysis, we can know that these three fitting parameters are independent, which ensures the uniqueness of the fitting curve. As previously reported, electric dipole p ≈ $3\times10^{-8}$ C m and the local electrostatic field $P_0/\varepsilon \approx 1$ V $nm^{-1}$ for 300nm P(VDF-TrFE) polymer,[14] the calculated dimensionless parameter k ≈ 1.19 which is consistent with the fitting one.

**CONCLUSIONS**

In summary, it is demonstrated that PL intensity of intralayer excitons in bilayer $WSe_2$ can be modulated (suppressed or enhanced) reversibly utilizing the built-in electric field from polarized P(VDF-TrFE) polymer thin film. The observed PL change is attributed to the intervally electrons transfer from two conduction band extreme (K and $\Lambda$ point). Since the energy difference is only ~ 50 meV between K and $\Lambda$ points in the conduction band, this electron transfer can happen relative easily. This further changes the two competing carrier recombination paths (direct and indirect excitons) and modifies overall optical radiation efficiency. To further understand and quantitatively verify the proposed theory, a semi-analytical model is developed based on the transferred electron effect. The theoretical result matches the experimental result well. This work may pave the way for electrically tunable photoluminescence in multilayer TMDCs.

**Method**

**PL Measurement**. The PL signals excited by a 514.5 nm line of an Ar-ion laser were collected by tri-vista Raman spectroscopy equipped with a liquid nitrogen-cooled Si-CCD camera. A X 50 telephoto objective lens with a numerical aperture of 0.45 were used, and a laser spot size was ~2 μm in diameter. Relatively low excitation power (<100 μW) is used to reduce the thermal effect. After measuring the polarization up state (P+), the sample is annealed at 400 K to eliminate the remnant polarization of P (VDF-TrFE), and then the test under polarization down state (P-).

**Fabrication**. The bilayer WSe$_2$ was prepared by mechanical exfoliation from 2H-WSe$_2$ bulk crystal and transferred to a silicon substrate covered with 300nm thick silicon oxide. The source and drain of the device were fabricated by electron beam lithography, and then proceeded Pd/Au (15nm /45 nm) deposition and lift-off steps. Next, the device was annealed at 200 °C in an argon atmosphere for two hours. Then the P(VDF-TrFE) polymer solution with a molar ratio of 70%/30% was coated on the top of the device. The device is baked at 115 °C for 15 min. Finally, aluminum (10 nm) was deposited on the channel covered by P(VDF-TrFE) polymer by electron beam evaporation.

**DFT calculations.** The electronic structures of bilayer WSe$_2$ were calculated through the density functional theory (DFT) code VASP. In our calculation, the exchange correlation potential was dealt with hybrid functionals at both Perdew−Burke−Ernzerhof (PBE) levels,[34] and the cutoff energy was set to 450 eV. A vacuum of 20 Å perpendicular to the surface was used to demonstrate finite layer, and a 15 × 15 k-meshed grid was employed to describe the periodic properties of the bilayer WSe$_2$. The convergence criterion was restricted to less than $10^{-5}$ eV in energy. The atomic positions were fully relaxed until the Hellmann−Feynman forces acting on each atom were less than 0.01 eV Å$^{-1}$.

**ASSOCIATED CONTENT**

**Supporting Information:**

Details of the semi-analytical model based on the transferred electron effect; Back-gate dependent PL measurement; Temperature dependence of PL measurements;

**AUTHOR INFORMATION**

**Corresponding Authors**

*Email: (Z. Gong) gongzr@szu.edu.cn

*Email: (Q. Qin) qi.qin@szu.edu.cn

*Email: (X. Wang) xjwang@mail.sitp.ac.cn

ORCID

Sixin Zhu: 0000-0002-4263-6838

Xingjun Wang: 0000-0002-4833-4339

Zhirui Gong: 0000-0002-4626-797X

**Notes**

The authors declare no competing financial interest.

**ACKNOWLEDGMENTS**

The authors are grateful for the financial support provided by the National Natural Science Foundation of China (Grant No. 11874377), Natural Science Foundation of Shanghai (Grant No.18ZR1445700) and the Natural Science Foundation of Guang-dong Province (Grant No. 2019A1515011400).

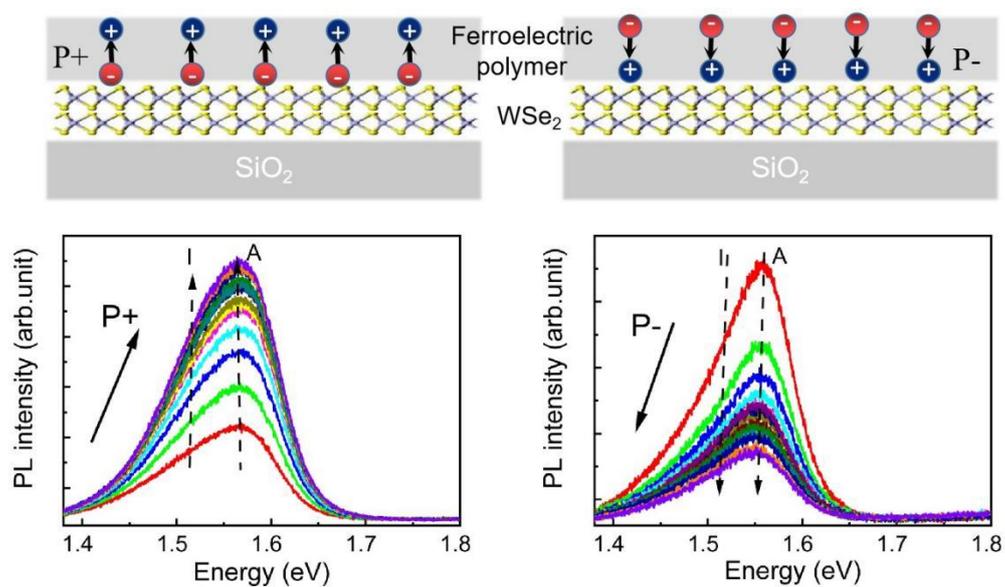

TOC graphic